\documentclass[pra,twocolumn,superscriptaddress]{revtex4-1}

\usepackage[T1]{fontenc}
\usepackage[latin9]{inputenc}
\usepackage[english]{babel} 
\usepackage{graphicx}
\usepackage{enumerate}
\usepackage{hyperref}
\usepackage{textcomp}
\usepackage{amsmath}
\usepackage{amssymb}
\usepackage{pgf}
\usepackage{diagbox}
\usepackage{booktabs}
\usepackage{bbold}
\usepackage{color}
\usepackage{bbm}
\usepackage{soul}
\usepackage{algorithm}
\usepackage{algpseudocode}
\usepackage{placeins}

\def\ud{\mathrm{d}}

\newcommand{\AC}{\mathrm{AC}}

\newcommand{\id}{\mathbb{1}}

\newcommand{\ini}{\mathrm{i}}
\newcommand{\fin}{\mathrm{f}}
\newcommand{\Ttrans}{\hat{\mathrm{T}}}

\begin{document}

\title{Superfluid to Mott transition in a Bose-Hubbard ring:\\ Persistent currents and defect formation}

\author{L. Kohn}
\affiliation{Scuola Internazionale Studi Superiori Avanzati (SISSA), Via Bonomea 265, 34136 Trieste, Italy}

\author{P. Silvi}
\affiliation{Center for Quantum Physics, and Institute for Experimental Physics, University of Innsbruck
		\&
		Institute for Quantum Optics and Quantum Information, Austrian Academy of Sciences, A-6020 Innsbruck, Austria}
	
\author{M. Gerster}
\affiliation{Institute for Complex Quantum Systems \& Center for Integrated Quantum Science and Technology (IQST), Ulm University, Albert-Einstein-Allee 11, D-89069 Ulm, Germany}

\author{M. Keck}
\affiliation{NEST, Scuola Normale Superiore \& Istituto Nanoscienze-CNR, I-56126 Pisa, Italy}

\author{R. Fazio}
\affiliation{Abdus Salam ICTP, Strada Costiera 11, I-34151 Trieste, Italy}
\affiliation{Dipartimento di Fisica, Universit\`a di Napoli ``Federico II'', Monte S. Angelo, I-80126 Napoli, Italy}

\author{G.E. Santoro}
\affiliation{Scuola Internazionale Studi Superiori Avanzati (SISSA), Via Bonomea 265, 34136 Trieste, Italy}
\affiliation{Abdus Salam ICTP, Strada Costiera 11, I-34151 Trieste, Italy}
\affiliation{CNR-IOM Democritos National Simulation Center, Via Bonomea 265, 34136 Trieste, Italy}

\author{S. Montangero}
\affiliation{Dipartimento di Fisica e Astronomia ``G. Galilei'', Universit\`a di Padova, I-35131 Padova, Italy}
\affiliation{Istituto Nazionale di Fisica Nucleare (INFN), Sezione di Padova, I-35131 Padova, Italy}
%\affiliation{Dipartimento di Fisica e Astronomia ``G.Galilei'', Universit\`a degli Studi di Padova \& INFN, I-35131 Padova, Italy}

\begin{abstract}
We revisit here the Kibble-Zurek mechanism for superfluid bosons  slowly driven across the transition towards the Mott-insulating phase. 
By means of a combination of the Time-Dependent Variational Principle and a Tree-Tensor Network, we characterize the current flowing during annealing in a ring-shaped 
one-dimensional Bose-Hubbard model with artificial classical gauge field on up to 32 lattice sites. 
We find that the superfluid current shows, after an initial decrease, persistent oscillations which survive even when the system is well inside the Mott insulating phase. 
We demonstrate that the amplitude of such oscillations is connected to the residual energy, characterizing the creation of defects while crossing the quantum critical point, 
while their frequency matches the spectral gap in the Mott insulating phase. Our predictions can be verified in future atomtronics experiments with neutral atoms in ring shaped traps. We believe that the proposed setup provides an interesting but simple platform to study the non-equilibrium quantum dynamics of persistent currents experimentally.
\end{abstract}
%\pacs{}
\maketitle

%\paragraph*{Introduction ---} 
\section{Introduction}
The Kibble-Zurek (KZ) mechanism, introduced to  understand  the defect formation at 
a symmetry-breaking phase transition, has been theoretically studied and experimentally verified in many different circumstances~\cite{PhysRep_Kibble_1980, PRL_Zurek_1997, Laguna_Zurek_PRD, PRL_Zurek_1998, PRL_Dziarmaga_1999, PRL_Fischer_2007, PRL_Dziarmaga_2008, PRL_Plenio_2010, PRL_Witkowska_2011, Science_Zurek_2012, WS_Campo_2014, PRL_Silvi_2016, PRA_Weiss_2018, Science_Chuang_1991, Science_Bowick_1994, Nature_Sadler_2006, PRB_Monaco_2009}. 
The KZ dynamics, initially investigated in classical (finite temperature) critical phenomena has been further extended to quantum critical systems, 
where it has also notable connections with many-body state preparation and 
adiabatic quantum computation/quantum annealing~\cite{Kadowaki_PRE98, Santoro_SCI02, PRB_Polkovnikov_2005, PRL_Zurek_2005, PRL_Dziarmaga_2005, Damski_2005, Lamacraft_PRL_2007, Franco_XXZ_KZ, DelCampo_PRL_2012, Chapman_QKZ_2016}. 
Progress in the realization of quantum simulators has led to important experimental results in KZ physics, most notably with cold atoms~\cite{Nature_Weiler_2008, 
Nature_Lamporesi_2013, Science_Navon_2015, Nature_Liu_2018, Nature_Keesling_2019} and trapped ions \cite{Nature_Pyka_2013, Nature_Ulm_2013}. 
Despite the multitude of works, the initial proposal put forward more than 30 years ago~\cite{Nature_Zurek_1985} --- to observe the formation of defects through the changes of 
the persistent current in a superfluid ring --- was never theoretically or experimentally addressed in quantum systems. The reason is twofold: Theoretically, the simulation of the dynamics of a quantum many-body system on a ring is a highly demanding task and has not been carried out so far. 
Experimentally, to date it has been difficult to realize a condensate with a ring-shape geometry.

\begin{figure}[t]
	\centering
	\begin{minipage}{3.4cm}
		\includegraphics[width=3.4cm]{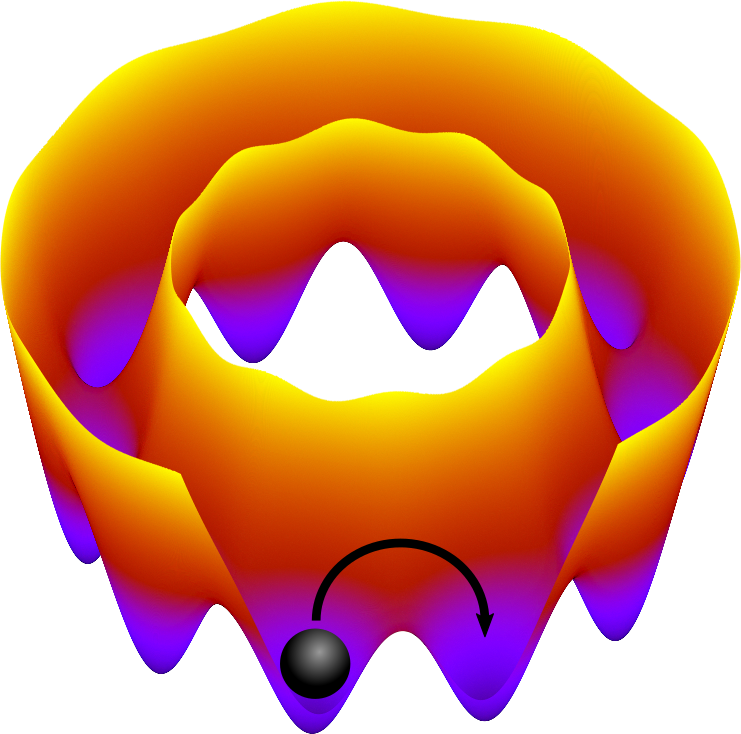}
	\end{minipage}
	\begin{minipage}{5.1cm}
		\includegraphics[width=5.15cm]{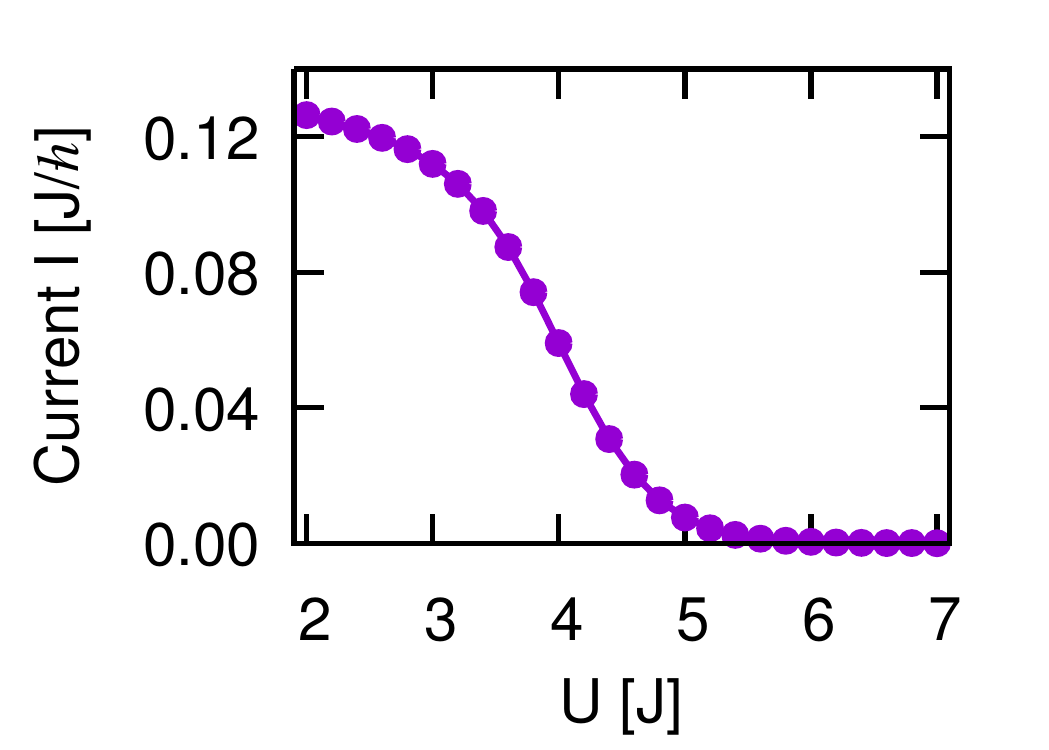}
	\end{minipage}
	\begin{picture}(0,0)
	\put(-87,35){$-Je^{i\phi/L}$}
	\end{picture}
	\caption{(Left) Schematic picture of the trapping potential. The particles are confined to lattice sites on a ring, where they can hop between neighboring
		sites with hopping amplitude $J$, picking up the phase $\pm \phi/L$.   
		(Right) Ground state current $I$, Eq.~\eqref{eqn:current_operator}, as a function of the on-site interaction $U$ in a system of 
		$L=32$ sites at unit filling for $\phi=0.7\pi$.}
	\label{fig:model}
\end{figure}
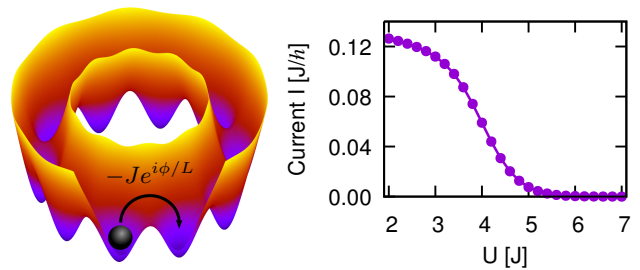

In this paper we fill this gap theoretically, by investigating how persistent currents are modified on crossing the superfluid-insulator transition in a Bose-Hubbard ring. 
Besides its conceptual interest, we believe that our work is timely in view of the experimental possibilities offered by the newly born field of atomtronics 
(see, e.g., the focus issue~\cite{NJP_Amico_2017}).  
We consider the case where a toroidal trapping potential and a lattice modulation along the trapping ring is present, such that the system 
is described by the Bose-Hubbard model on a ring pierced by an external static gauge field (see Fig.~\ref{fig:model}). 
While the ground state of this system has been investigated extensively~\cite{PRL_Cominotti_2014,PRA_Arwas_2017}, previous studies of the non-equilibrium time-evolution focused on the mean-field regime (e.g. large particle densities), chain geometries without persistent current, or short time analysis \cite{PRL_Fischer_2006, PRA_Fischer_2008, PRA_Cucchietti_2007, PRA_Jaksch_2004, PRA_Kollath_2012, PRA_Tuchman_2006, PRL_Natu_2011, PRA_Navez_2010}.
We present fully quantum, time-dependent results for the persistent current on a ring of up to $L=32$ sites, hence beyond the reach of exact diagonalization. 
We carry out our calculations using an approach which combines the Time-Dependent Variational Principle (TDVP)~\cite{PRL_Haegeman_2011,PRB_Haegeman_2016} 
with Tree-Tensor Networks (TTN)~\cite{PRA_Shi_2006,PRA_Silvi_2010, PRB_Gerster_2014,montangero2018introduction,SciPost_Silvi_2019}. 
We show that the annealing from superfluid to a Mott insulator --- which does not carry any current itself at equilibrium --- leads to oscillations of the persistent 
current flowing in the ring, with the amplitude showing a strong dependence on the annealing rate and being related to the residual energy, 
a key quantity in the KZ mechanism. 
We also show that the frequency of the persistent current oscillations matches the spectral gap of the Mott insulating final state. 

Induced by a classical gauge field, the magnitude of the persistent current of bosonic particles flowing on a ring strongly depends on the interaction between 
particles. At equilibrium, strong repulsion at integer filling typically leads to Mott localization, suppressing the persistent current, while weak interactions allow current
flow in presence of an external field~\cite{PRL_Cominotti_2014}. 
Persistent currents are of particular interest in the growing field of atomtronics~\cite{PRL_Ramanathan_2011, PRL_Wright_2013, PRL_Labouvie_2015, QST_Haug_2018}, where atomic particle currents are used to mimic currents in electronic devices~\cite{PRA_Seaman_2007}.
For instance, atomtronic quantum interference devices  (AQUIDs), being the analogue of superconductor based SQUIDs, 
allow to study persistent currents in highly controllable systems of ultracold atomic gases~\cite{Nature_Greiner_2002, Nature_Trotzky_2012}. In these systems 
of neutral atoms, currents are created by confining a Bose-Einstein condensate to an effectively one-dimensional system and driving the particles through gauge fields, implemented either artificially or by stirring, e.g. using a rotating potential barrier~\cite{RMP_Dalibard_2011, Aidelsburger_PRL_2011}. Thanks to the unrivaled tunability of 
interactions and potentials in combination with low decoherence, these systems provide experimental tools to study new collective phenomena, with possible 
applications in the development of high precision sensors, quantum simulation or quantum information processing~\cite{NJP_Amico_2017}, where for instance the 
superposition of persistent current states could serve as an implementation of qubits~\cite{Nature_Amico_2014,NJP_Aghamalyan_2015}. 

The paper is organized as follows: In Sec.~\ref{sec:model}, we define the time-dependent Bose-Hubbard model considered throughout the paper and discuss some of its equilibrium features. Sec.~\ref{sec:TTN} is dedicated to our Tree-Tensor Network algorithm. We briefly recapitulate the Time-Dependent Variational Principle and present the algorithm to carry out the time evolution of a TTN. Results are presented in Sec.~\ref{sec:results}, where we discuss the behavior of the persistent current, its connection to the residual energy, and the system size dependence of our findings. In Sec.~\ref{sec:perturbation_} we provide details of our semi-analytical calculations for the oscillation frequency of the current using second order perturbation theory. We finally conclude the paper in Sec.~\ref{sec:conclusions} with some additional remarks on our results and possible experimental realizations.

%\paragraph*{The model ---} 
\section{Time-dependent Bose-Hubbard model on a ring} \label{sec:model}
For a system consisting of $L$ sites the Bose-Hubbard model, pierced by a magnetic field, is described by the Hamiltonian
\begin{equation} 
\label{eqn:HsysF}
	\hat{H} = -J\sum\limits_{j=1}^{L} \left( e^{i\phi/L }b_{j+1}^{\dagger}b_{j} + H.c. \right) +\frac{U}{2}\sum\limits_{j=1}^{L}n_j(n_j-1),
\end{equation}
where $J$ and $U$ are the hopping amplitude and on-site interaction, respectively, and the Peierls' phase $\phi$ takes into account the  
flux $\Phi$ through the ring in units of the flux quantum $\Phi_0$ ($\phi=2\pi \, \Phi/\Phi_0$). 
From now on, we will simply refer to $\phi$ as the ``magnetic field'', for simplicity.  
We work at fixed density of $\rho=1$ particle per site, where the model exhibits, for $\phi=0$, an equilibrium quantum phase transition between a Mott insulator 
and a superfluid at $(U/J)_c\approx 3.37$~\cite{PRB_Kuhner_2000}.  
Still at equilibrium, but for $\phi\neq 0$, based on previous mean-field studies~\cite{PRB_Oktel_2007} and on a strong coupling analysis of the 
two-dimensional case~\cite{PRB_Niemeyer_1999}, the critical value $(U/J)_c$ is expected to decrease as compared to the zero-field case, 
therefore extending the Mott insulating phase.

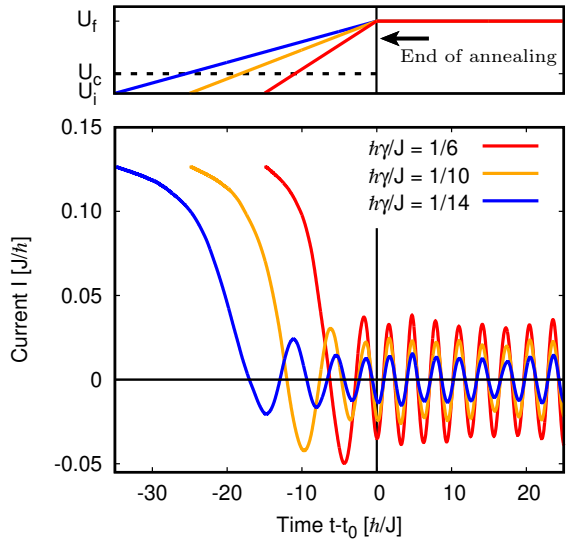
\begin{figure}[t]
	\centering
	\begin{picture}(0,0)
		\put(155,189){\scriptsize End of annealing}
	\end{picture}
	\includegraphics[width=8cm]{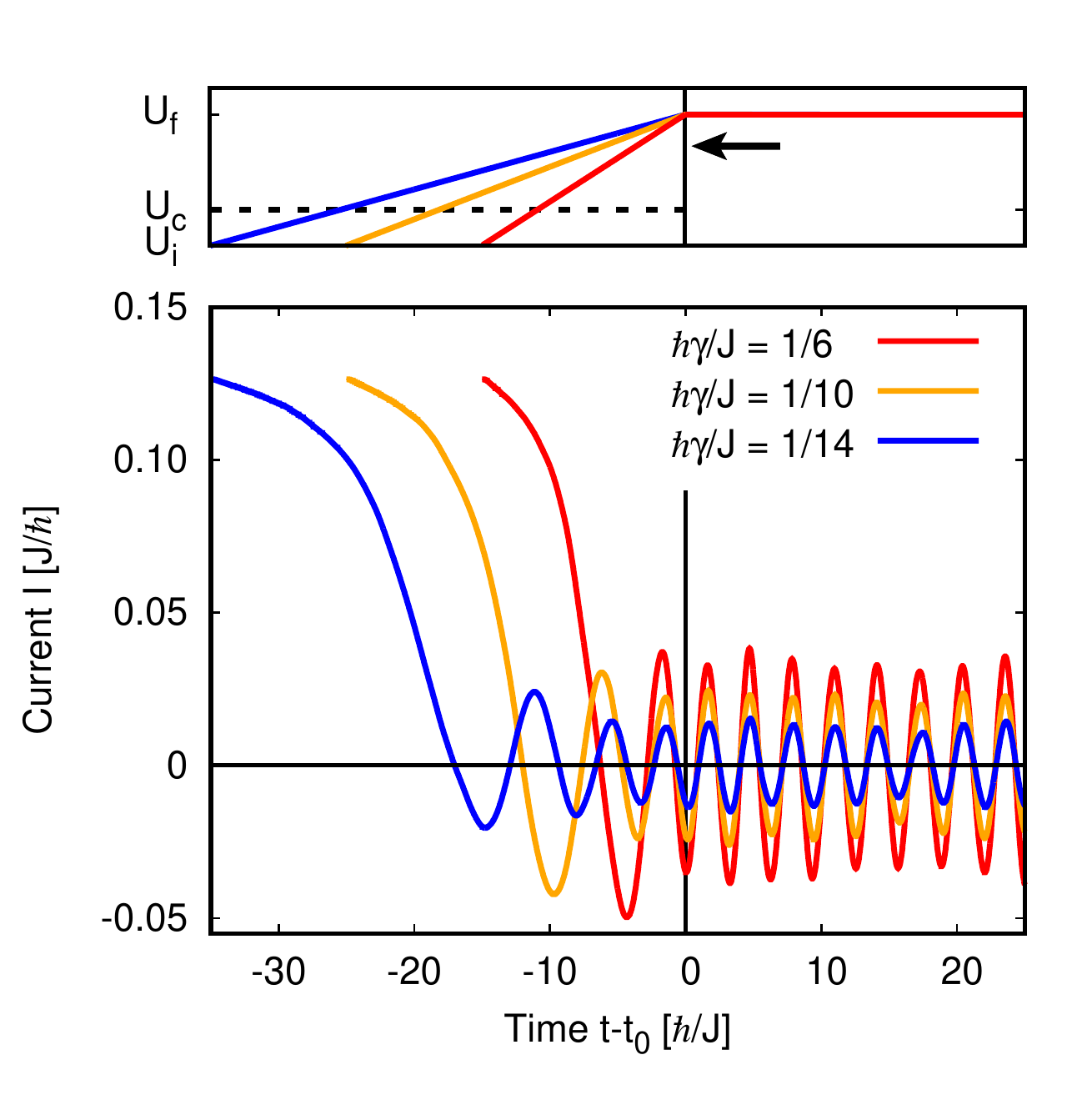}
	\caption{(Top) Annealing protocol used for the on-site interaction $U(t)$ throughout the paper. 
		After linearly increasing the interaction with annealing rate $\gamma$, the system for $t\ge t_0$ is evolved 
		according to the final constant Hamiltonian. 
		(Bottom) Current $I(t)$ as a function of time during the annealing from $U_\ini=2J$ to $U_{\fin}=7J$ in a system 
		of $L=32$ sites at fixed magnetic field $\phi=0.7\pi$ for three different annealing rates $\gamma$. }
	\label{fig:cur_gamma_ttn}
\end{figure}

We consider an out-of-equilibrium dynamics, in which the initial superfluid in the presence of a magnetic field $\phi\neq 0$ is driven in time across 
the transition towards the Mott insulating phase by ramping-up the Hubbard interaction $U$ (or, equivalently, by ramping down the hopping matrix 
element $J$). This procedure is the reverse as compared to the typical Kibble-Zurek scenario discussed in Ref.\cite{Nature_Zurek_1985}, where the transition from the disordered to the ordered phase is considered, corresponding to a Mott to superfluid transition for the Bose-Hubbard model. However, it has been shown in Ref.\cite{PRB_Dziarmaga_2017} that also the superfluid to Mott transition obeys Kibble-Zurek power law scaling within a limited range of annealing velocities. In contrast to the superfluid-Mott transition, excitations are created only after crossing the transition, meaning that there are almost no excitations within the superfluid phase. The  quantity we investigate is the time-dependent current flowing on the ring, defined as the expectation value 
$I(t) = \langle \psi(t) | \hat{I} | \psi(t)\rangle$  of the current operator
\begin{align} 
	\label{eqn:current_operator}
	\hat{I} 
	= -\frac{1}{\hbar} \frac{\partial \hat{H}}{\partial \phi} 
	= \frac{iJ}{\hbar L}\sum\limits_{j=1}^{L} \left( e^{i\phi/L }b_{j+1}^{\dagger}b_{j} - H.c. \right) \;.
\end{align}
More in detail, we prepare the system in the ground state $|\psi_0\rangle$ of the Hamiltonian $\hat{H}$ within the superfluid phase,
specifically for $U(t=0)=U_\ini = 2J$, with a given value of the external magnetic flux $\Phi< \Phi_0/2$, corresponding to $\phi=2\pi \Phi/\Phi_0$ 
in the interval $[0,\pi)$ --- kept fixed during the dynamics ---, and then anneal the value of the interaction $U(t)$ up towards 
a final value $U(t=t_0)=U_\fin$ deep inside the Mott phase, in a time $t_0$, at a constant rate $\gamma=(U_\fin/U_{\ini}-1)/t_0$. 
The ramp is followed by a final part of the evolution where the interaction is kept constant at $U_\fin$. 
The time-evolution of the interaction is therefore given by (see also Fig.~\ref{fig:cur_gamma_ttn}, top)
$U(t) = U_\ini(1 + \gamma t)$ for $t\le t_0$, while $U(t) = U_\fin$ for $t>t_0$.

%---------------------------------
\section{The TTN-TDVP algorithm} \label{sec:TTN}
%---------------------------------%

In this section we provide details on the time-evolution algorithm, the Time-Dependent Variational Principle (TDVP) applied to a Tree-Tensor Network (TTN). The TTN {\it Ansatz} is well-suited for one- and two-dimensional systems with periodic boundary conditions, since the distance between the first and last sites scales with only $\mathcal{O}(\log(L))$ in the Tensor Network~\cite{PRB_Tagliacozzo_2009, PRB_Gerster_2014,NJP_Gerster_2016,PRB_Gerster_2017}. In particular we employ a two-tensor evolution scheme, allowing to adapt the bond dimension during the simulation dynamically. Starting from the TDVP as introduced by Dirac and Frenkel, 
it has been formulated for loopless Tensor Networks by Haegeman \textit{et al.} \cite{PRL_Haegeman_2011}. Only recently this method has been further improved \cite{SIAM_Lubich_2015,PRB_Haegeman_2016}, overcoming problems with small singular values in the original formulation. The algorithm originally has been formulated for Matrix Product States (MPS) and has been extended for MPS with an optimized boson basis \cite{PRB_Schroeder_2016}, introducing additional tensors into the MPS.

In general, the idea of the TDVP algorithm for Tensor Networks is to project the change of the wave function onto the tangent space of the Tensor Network manifold $\mathcal{M}$ with given bond dimension $D$. This guarantees that each update of a tensor is optimal, meaning that the Euclidean distance between the exact evolution and its MPS approximation is minimal \cite{PRB_Haegeman_2016}. Formally the projection is introduced into  Schr{\"o}dinger's equation by means of the projection operator $\hat{P}_{|\psi(t)\rangle}$, projecting on the tangent space of $\mathcal{M}$ at $|\psi(t)\rangle \in \mathcal{M}$:
\begin{equation}
i\hbar \frac{\ud}{\ud t}|\psi(t)\rangle = \hat{P}_{|\psi(t)\rangle}\hat{H}|\psi(t)\rangle, \label{eq:TDVP}
\end{equation} 

For the model considered in this paper we exploit the U(1) symmetry of the Hamiltonian explicitly\cite{PRA_Singh_2010,PRB_Singh_2011,SciPost_Silvi_2019} and we use a two-tensor integration scheme in order to allow for a dynamical adjustment of the bond dimension and the dimensions of the different symmetry sectors. For MPS this two-tensor algorithm was presented in Ref.~\cite{PRB_Haegeman_2016}. It has been pointed out that TDVP cannot be formulated in the form of a differential equation as in Eq.~\eqref{eq:TDVP} for the two-tensor scheme, because the projector does not keep the state within the Tensor Network manifold. However, the algorithm requires a discrete time step anyway and therefore we can perform a Singular Value Decomposition (SVD) after each update to bring the Tensor Network back to the variational manifold $\mathcal{M}$. We present this two-tensor integration scheme for a TTN, which can easily be generalized to arbitrary loopless Tensor Networks.

\begin{figure}[t]
	\centering
	\includegraphics[width=8.45cm]{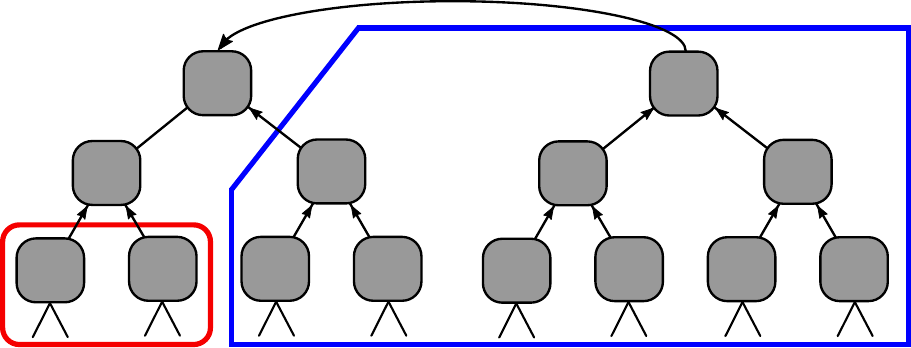}
	\begin{picture}(0,0)
	\put(-226,4){ $Q^{(\lambda)}$}
	\put(-227,42){ $C^{(\lambda)}_L$}
	\put(-198,65){ $C^{(\lambda)}_R$}
	\put(-130,60){ $U^{(\lambda)}$}
	\end{picture}
	\caption{Visualization of the Tree-Tensor Network decomposition in Eq.~\eqref{eq:TTNdecom} for a system with 16 sites. In this example the matrix $Q^{(\lambda)}$ is built from the two lower left tensors encircled in red. The center block is composed of the tensors $C^{(\lambda)}_L$ and $C^{(\lambda)}_R$, being the orthogonality center of the network as indicated by the arrows, while the matrix $U^{(\lambda)}$ represents all the rest of the network (blue box). Note that this notation is used for convenience to represent the  three different parts of the TTN: The orthogonal parts $Q^{(\lambda)}$ and $U^{(\lambda)}$, and the center block $C^{(\lambda)}$. In practice we never calculate $Q^{(\lambda)}$ or $U^{(\lambda)}$ explicitly but keep them in the TTN format.}
	\label{fig:Tree_decomp}
\end{figure}
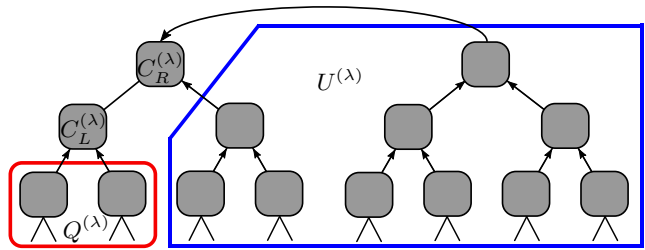

Let us introduce a decomposition of a Tree-Tensor Network consisting of $M+1$ tensors. Since the network is loopless, there are $M$ pairs of tensors, with the tensors of a pair being connected through a bond link. Then, for $\lambda=1,..,M$ labeling such a pair of tensors, as depicted in Fig.~\ref{fig:Tree_decomp}, the state represented by the network reads 
\begin{equation}
|\psi\rangle = \sum\limits_{k,m} \left(Q^{(\lambda)}C^{(\lambda)}(U^{(\lambda)})^{\dagger}\right)_{k,m} |\Phi^{(\lambda)}_{L,k}\rangle|\Phi^{(\lambda)}_{R,m}\rangle, \label{eq:TTNdecom}
\end{equation}
with $C^{(\lambda)}=C^{(\lambda)}_LC^{(\lambda)}_R$ being the matrix corresponding to the two-tensor center block: This matrix is obtained by multiplying the matrices corresponding to the tensors of pair $\lambda$, $C^{(\lambda)}_L$ and $C^{(\lambda)}_R$, where the tensors need be chosen such that in the network geometry $C^{(\lambda)}_R$ is closer to the pair $\lambda=M$ than $C^{(\lambda)}_L$. Note that for simplicity we use the same notation for tensors and the corresponding matrices formed by fusing tensor indices. The states $|\Phi^{(\lambda)}_{L,k}\rangle$ and $|\Phi^{(\lambda)}_{R,m}\rangle$ in Eq.~\eqref{eq:TTNdecom} are product states in the local basis. They correspond to the two subsystems obtained by splitting the network between the matrices $C^{(\lambda)}_L$ and $C^{(\lambda)}_R$. In Fig.~\ref{fig:Tree_decomp} we visualize this decomposition in an example. Note that even though the labels indicate a separation into a left and right part, the physical bipartition usually is of different shape. %Instead, left and right part of the network are defined through the property that the special pair of tensors labeled by $\lambda=M$ is never contained in the left part, represented by the matrix $Q^{(\lambda)}$.

Since the network is free of loops we can choose it to be isometrized with respect to the center block $C^{(\lambda)}$, such that $Q^{(\lambda)}$ and $U^{(\lambda)}$  have orthonormal columns $\left( (V^{(\lambda)})^{\dagger}V^{(\lambda)}=\id, V=Q,U\right)$. 
Using this decomposition it is possible to follow the derivations in  Refs.~[\onlinecite{PRB_Schroeder_2016,SIAM_Lubich_2015,SIAM_Lubich_2013}], to find the projector onto the space of two-tensor variations: 
\begin{equation*}
\hat{P}=\hat{P}^{(2)}_{M} + \sum\limits_{\lambda=1}^{M-1}\left(\hat{P}^{(2)}_{\lambda}-\hat{P}^{(1)}_{\lambda}\right) ,
\end{equation*}
with $\hat{P}^{(2)}_{\lambda}=\hat{P}_{L,\lambda}\otimes\hat{P}_{R,\lambda}$, $\hat{P}^{(1)}_{\lambda}=\hat{P}_{L+C,\lambda}\otimes\hat{P}_{R,\lambda}$ and

\begin{align*}
\hat{P}_{L,\lambda}&=\sum\limits_{k,k'} \left( Q^{(\lambda)}(Q^{(\lambda)})^{\dagger} \right)_{k,k'} |\Phi^{(\lambda)}_{L,k}\rangle\langle \Phi^{(\lambda)}_{L,k'} | \\
\hat{P}_{L+C,\lambda}&=\sum\limits_{k,k'} \left( Q^{(\lambda)}C_{L}^{(\lambda)}(C_{L}^{(\lambda)})^{\dagger}(Q^{(\lambda)})^{\dagger} \right)_{k,k'} |\Phi^{(\lambda)}_{L,k}\rangle\langle \Phi^{(\lambda)}_{L,k'} | \\
\hat{P}_{R,\lambda}&=\sum\limits_{m,m'} \left( U^{(\lambda)}(U^{(\lambda)})^{\dagger} \right)_{m,m'} |\Phi^{(\lambda)}_{R,m}\rangle\langle \Phi^{(\lambda)}_{R,m'}|.
\end{align*}

\begin{figure}[h]
	\begin{minipage}{0.48\textwidth}
		\includegraphics[scale=0.54]{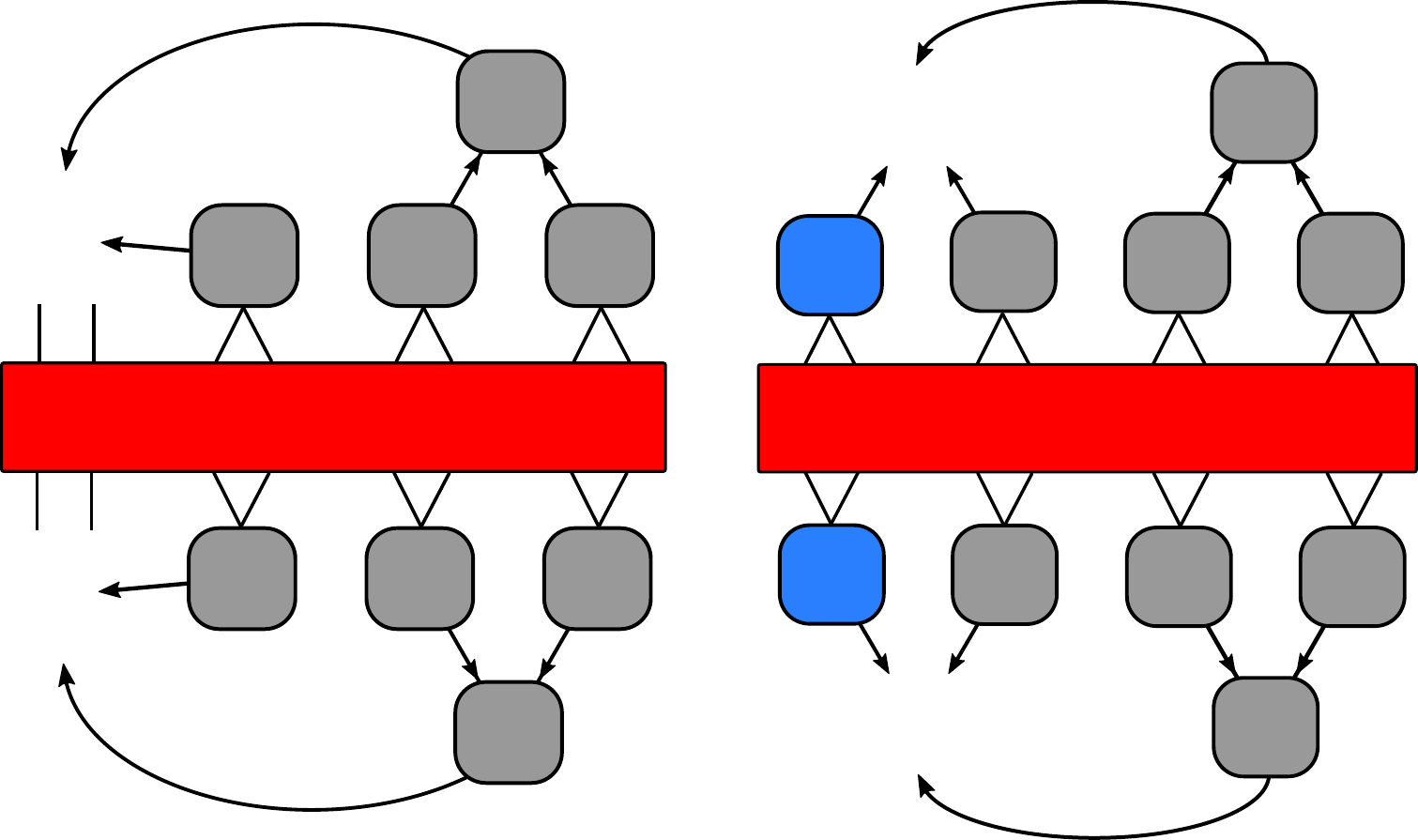}
	\end{minipage}
	\caption{(Left) Effective Hamiltonian for the two-tensor block $C^{(1)}$ in a TTN of 8 physical sites. It is obtained by contracting the Hamiltonian (red rectangle) with the state from above and below. The state is isometrized with respect to the tensor $C^{(1)}$, taken out from the state. (Right) Effective Hamiltonian for the tensor $C^{(1)}_R$ built from the contraction of the effective Hamiltonian on the left with the time-evolved tensor $C^{(1)}_L$ from above and below. }
	\label{fig:eff_ham}
\end{figure}

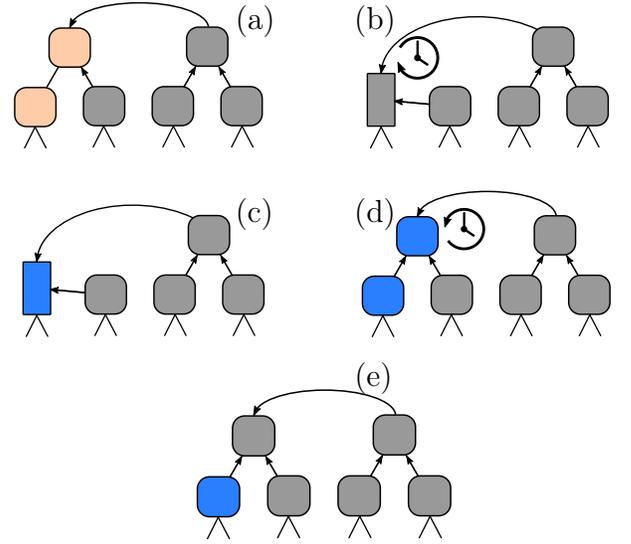
\begin{figure}
	\includegraphics[scale=0.8]{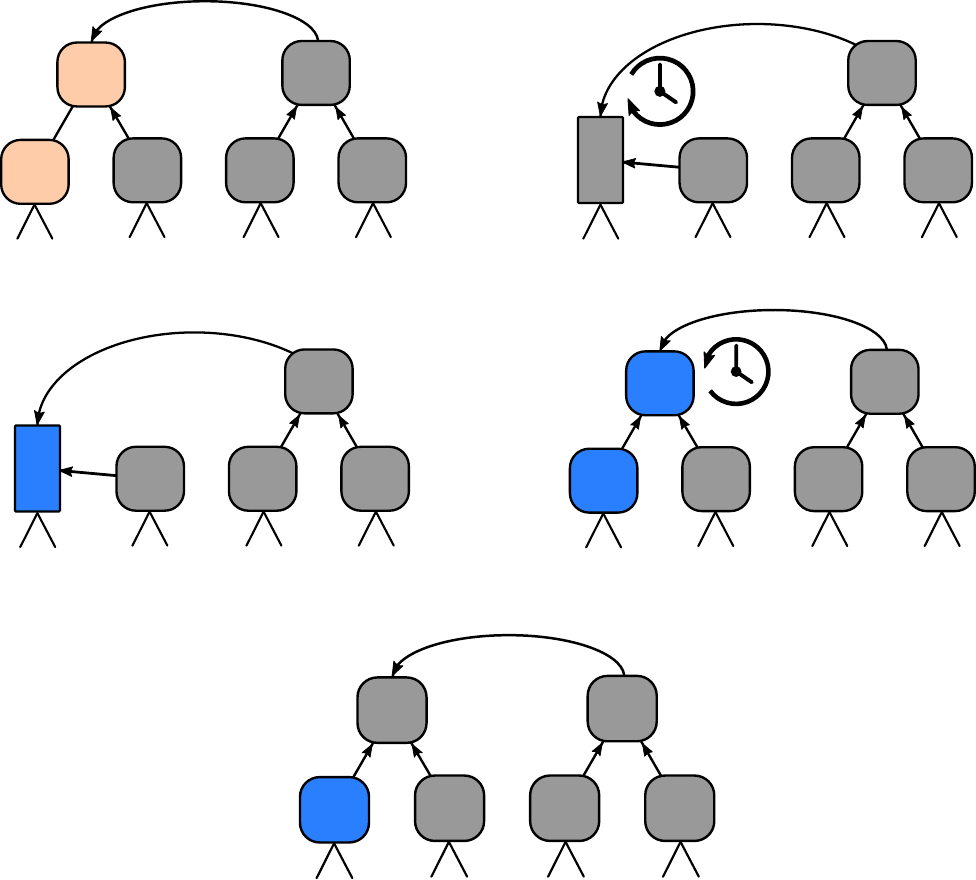}
	\large
	\begin{picture}(0,0)
	\put(-145,193){(a)}
	\put(-100,193){(b)}
	\put(-145,120){(c)}
	\put(-100,120){(d)}
	\put(-100,58){(e)}
	\end{picture}
	\caption{Pictorial description of the algorithm to evolve the block built out of the two leftmost (orange) tensors by one time step, in a system of 8 sites. (a) Initially the TTN is isometrized with respect to the orange tensors building the block (isometrization indicated by arrows) (b) The two tensors are contracted to form a single block, being evolved according to Eq.~\eqref{eq:update_tens1} with the effective Hamiltonian in Fig.~\ref{fig:eff_ham}(left). (c) The blue color indicates that the block has been evolved in time. Afterwards, the block is split by means of an SVD. The singular values are contracted into the "right" tensor, being the new orthogonality center of the network. (d) The "right" tensor is evolved backwards in time according to Eq.~\eqref{eq:update_tens2}, using the effective Hamiltonian in Fig.~\ref{fig:eff_ham}(right). (e) The first pair of tensors has been evolved by one time step. The upper tensor of the evolved pair is drawn in grey again as it has been evolved backwards in time.}
	\label{fig:evolve}	
\end{figure}

One of the key ideas of the TDVP in this form is to use a Lie-Trotter splitting \cite{Springer_Hairer_2006} for Eq.~\eqref{eq:TDVP}, yielding one differential equation for each of the summands in the projector. These differential equations can be solved efficiently one after the other \onlinecite{SIAM_Lubich_2015}. We can therefore update the network according to every summand in the projector step by step. Since there are two types of projectors, $\hat{P}^{(2)}_{\lambda}$ and $\hat{P}^{(1)}_{\lambda}$, we get two structurally different equations to update the network. In particular the projector  $\hat{P}^{(2)}_{\lambda}$ introduces the update of a two-tensor block $C^{(\lambda)}$, while the projector $\hat{P}^{(1)}_{\lambda}$ leads to an update of the tensor $C^{(\lambda)}_R$. Effectively these updates are governed by the equations

\begin{align}
i\hbar\dot{C}^{(\lambda)}&= \ \ \mathcal{H}^{(2)}_{\text{eff},\lambda}C^{(\lambda)} \qquad \  \lambda=1,..,M \label{eq:update_tens1} \\ 
i\hbar\dot{C}^{(\lambda)}_R&=-\mathcal{H}^{(1)}_{\text{eff},\lambda}C^{(\lambda)}_R \qquad \lambda=1,..,M-1,
\label{eq:update_tens2}
\end{align}
where the application of the effective Hamiltonian on the right hand side corresponds to a tensor contraction of multiple indices. The effective Hamiltonians are constructed from the summands on the right hand side of Eq.~\eqref{eq:TDVP} as depicted in Fig.~\ref{fig:eff_ham} for the simple case of a two-layer tree (8 physical sites), where the center block is built from the left lower and upper tensors. Note that after evolving the two-tensor block an SVD has to be performed in order to bring the network back to the initial form. The full algorithm is summarized in Alg.~\ref{alg:ttn_tdvp}, while some of the steps are shown graphically in Fig.\ref{fig:evolve}. In our implementation we chose the pair built from the uppermost tensors to be the one with $\lambda=M$, for which there is no backwards evolution. The tensors in the TTN are then evolved in time from left to right, bottom to top.

\begin{algorithm}[H]
	\caption{2-tensor TTN-TDVP}
	\label{alg:ttn_tdvp}
	\begin{algorithmic}[1]
		\For {$\lambda =1,..,M$}
		\State Isometrize network w.r.t. $C^{(\lambda)}$
		\State Build $\mathcal{H}^{(2)}_{\text{eff},\lambda}$
		\State Evolve $C^{(\lambda)}$ according to Eq.~\eqref{eq:update_tens1}
		\State Perform SVD of
		$C^{(\lambda)}=C^{(\lambda)}_LC^{(\lambda)}_R$ $\left( (C_{L}^{(\lambda)})^{\dagger}C_{L}^{(\lambda)}=\id \right) $
		\If {$\lambda\ne M$}
		\State Build $\mathcal{H}^{(1)}_{\text{eff},\lambda}$
		\State Evolve $C^{(\lambda)}_R$ according to Eq.~\eqref{eq:update_tens2}
		\EndIf
		\EndFor
	\end{algorithmic}
\end{algorithm}

\section{Results} \label{sec:results}  
Since the system starts in the ground state $|\psi_0\rangle$ at a finite value of the external flux $\phi\in [0,\pi)$ and inside 
the superfluid phase, it  displays an initial  persistent current $I(0)=\langle \psi_0 | \hat{I} | \psi_0\rangle$ (see Fig.~\ref{fig:model}). 
Upon annealing the interaction upwards to enter in the Mott region of the equilibrium phase diagram, the current $I(t)$ drops  towards zero (see Fig.~\ref{fig:cur_gamma_ttn}), 
the expected value of the equilibrium current at zero temperature in the Mott phase. 
There is however a residual, time-dependent current which is approximately sinusoidal, $I_{\AC}(t) \sim I_0 \cos(\omega_0 t + \varphi)$, with a characteristic 
main frequency $\omega_0$. 
The amplitude $I_0$  depends on the annealing rate $\gamma$ and will be the key signal of the defects created by crossing the quantum phase 
transition: The slower is the annealing, the smaller is $I_0$, as suggested by the three curves in Fig.~\ref{fig:cur_gamma_ttn}. 
As expected, this extrapolates well to the adiabatic limit ($\gamma \rightarrow 0$), where no oscillations are present, since the equilibrium current of the Mott 
insulator vanishes. 
On the other hand, the main frequency $\omega_0$ is essentially unaffected by the annealing rate, but increases when the final Mott interaction $U_{\fin}$ 
is increased. As we will show, this behavior is related to the dynamical spectral gap of the final Mott insulating state, as the persistent alternating current is related 
to the dynamical oscillation between the Mott insulating state and higher excited states in which few holes and multiply occupied sites are created.
 
In the following, we analyze the current amplitude $I_0$ and its relation to the residual energy $\epsilon_{\text{res}}$, defined as the excitation energy above the final ground state energy, and the oscillation frequency $\omega_0$ 
in more detail. At the end of the section we will discuss the size dependence of the results.

\begin{figure}[t]
	\centering
	
	\includegraphics[width=8.5cm]{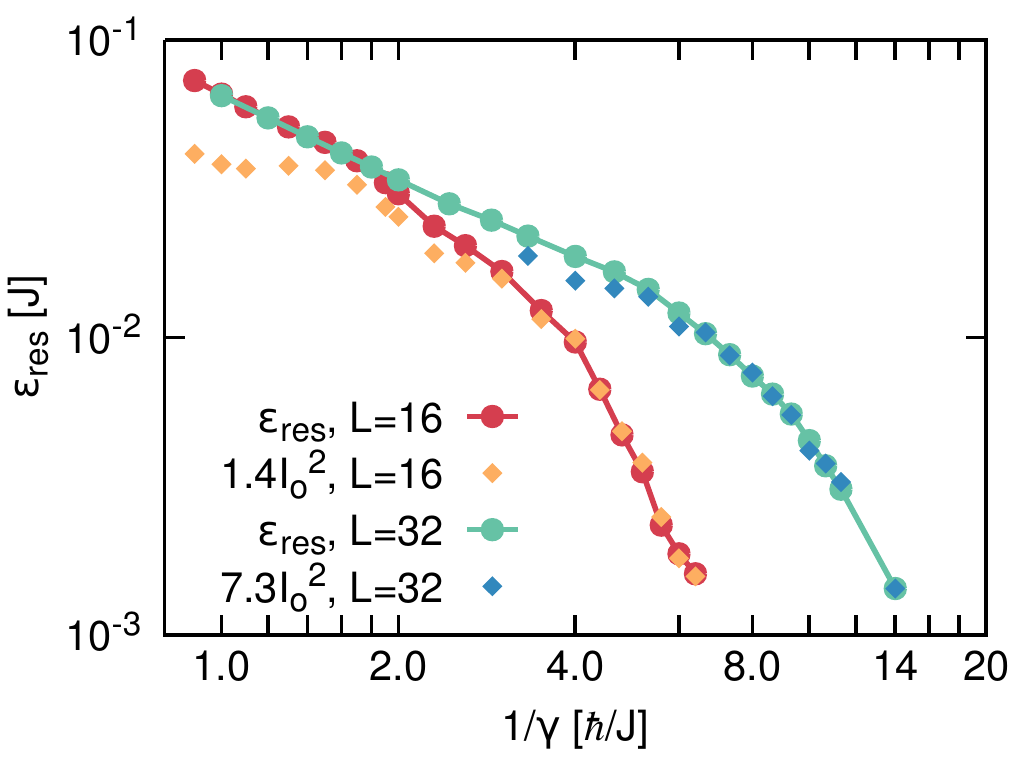}
	\caption{ Residual energy $\epsilon_{\text{res}}$ and square of the oscillation amplitude $I_0^2$ of the persistent current as a function of the inverse annealing rate $\gamma$ in systems of sizes $L=16,32$ at fixed field $\phi=0.7\pi$. The on-site interaction is ramped from $U_\ini=2J$ to $U_\fin=7J$.}
	\label{fig:res_energy_gamma_ttn}
\end{figure}

We extract the oscillation amplitude by numerically calculating $I_0 = (1/2)[\max\limits_{t\ge t_1}I(t) -  \min\limits_{t\ge t_1}I(t) ]$, where we take $t_1>t_0$ in 
order to neglect some possible transient behavior after the final value of $U_{\fin}$ is reached. The results are presented in 
Fig.~\ref{fig:res_energy_gamma_ttn}, quantifying the decay of the amplitude with decreasing annealing rate $\gamma$. 
As shown hereafter, this behavior is understood by the dependence of the amplitude on the occupation of higher excited states, and in particular the occupation $c_2$ of the first excited state of the final Hamiltonian. The probability $c_2$, on the other hand, behaves like in a two-level system undergoing a Landau-Zener dynamics, with decreasing excitation probability as $\gamma\rightarrow0$. 
The amplitude $I_0$  is the main signature of the quasi-adiabatic driving through the transition. This is further corroborated by a relation between the current amplitude and the residual energy as detailed below.
Denoting by $\{|\alpha\rangle\}$ the (many-body) eigenstates of the final Hamiltonian $\hat{H}_{\fin}$, with associated energies $E_{\alpha}$, 
one can write for any time $t>t_0$:
\begin{equation}
I(t) = \sum_{\alpha,\alpha'} c_{\alpha'}^{\star} c_{\alpha} e^{-i(E_{\alpha}-E_{\alpha'})(t-t_0)/\hbar}
\langle \alpha' | \hat{I} | \alpha \rangle \;.
\label{eq:cur_decom}
\end{equation}
The constants $c_{\alpha}=\langle \alpha | \psi(t_0) \rangle$ are the overlaps between the eigenstates of $\hat{H}_{\fin}$ and the state $|\psi(t_0)\rangle$ at the 
end of the annealing ramp. The diagonal matrix elements of the current operator can lead to a DC-component of the current, while off-diagonal elements can cause 
oscillations of the current. Let us now assume that the annealing ramp is slow, such that the system mainly remains in the ground state $|1\rangle$ and only 
one excited state $|2\rangle$, compatible with the translational symmetry, is slightly occupied during the ramp $(|c_2|\ll |c_1|\Rightarrow c_1=1+\mathcal{O}(c_2^2))$, 
while all the other excited states have negligible occupation. 
Then Eq.~\eqref{eq:cur_decom} simplifies and becomes
$I(t) \sim |c_2 \langle 1 | \hat{I} | 2 \rangle | \cos(\Delta \cdot t+\varphi) +I_{\text{offset}}$, 
where $\varphi$ is an unimportant phase, while $\hbar\Delta=E_2-E_1$ is the energy gap between ground and first excited state.
Here $I_{\text{offset}}$ is a DC-component, dominated by the ground state contribution, and vanishing deep in the Mott phase, as 
demonstrated in Fig.~\ref{fig:model}. 
On the other hand we can argue in the same way that the residual energy will be dominated by the contribution of the 
first excited state $|2\rangle$:
\begin{align}
\epsilon_{\text{res}}\approx|c_2|^2\hbar\Delta \;.
\end{align}
In the slow annealing regime, this implies that both the current amplitude ($I_0\propto |c_2|$) and the residual energy ($\epsilon_{\text{res}} \propto |c_2|^2$) depend on the annealing rate $\gamma$ only through the occupation $c_2(\gamma)$ of the first excited state. 
In Fig.~\ref{fig:res_energy_gamma_ttn} this relation ($\epsilon_{\text{res}}(\gamma) \propto I_0^2(\gamma)$) is verified numerically for different $\gamma$, 
by comparing the residual energy to the square of the current (multiplied by a size dependent prefactor).
Clearly, the agreement is quite good, in the slow annealing regime. 
Theoretically the propotionality prefactor is given by $\epsilon_{\text{res}}/I_0^2=\hbar\Delta/|\langle 1 | \hat{I} | 2 \rangle|^2$, and has been obtained by 
fitting this ratio for slow annealing ramps to a constant. 

As explained in Ref.\cite{PRB_Dziarmaga_2017} the residual energy shows an effective power-law behavior within a limited range of annealing velocities, which we find for $1/\gamma \lesssim 3$ (L=32) and $1/\gamma \lesssim 1.7$ ($L=16$). However, in this regime the simple picture of having only one excited state involved in the dynamics breaks down, as manifested by the disagreement of the current and the residual energy for $L=16$ and slightly indicated by the latest points for $L=32$. Very fast annealing leads to a rapid increase of entanglement, preventing us from studying this regime for large systems. Moreover, fast annealing leads to the appearance of additional frequencies in the current, with a more quantitative analysis, however, being beyond the scope of this work.

\begin{figure}[t]
	\centering
	\includegraphics[width=8.7cm]{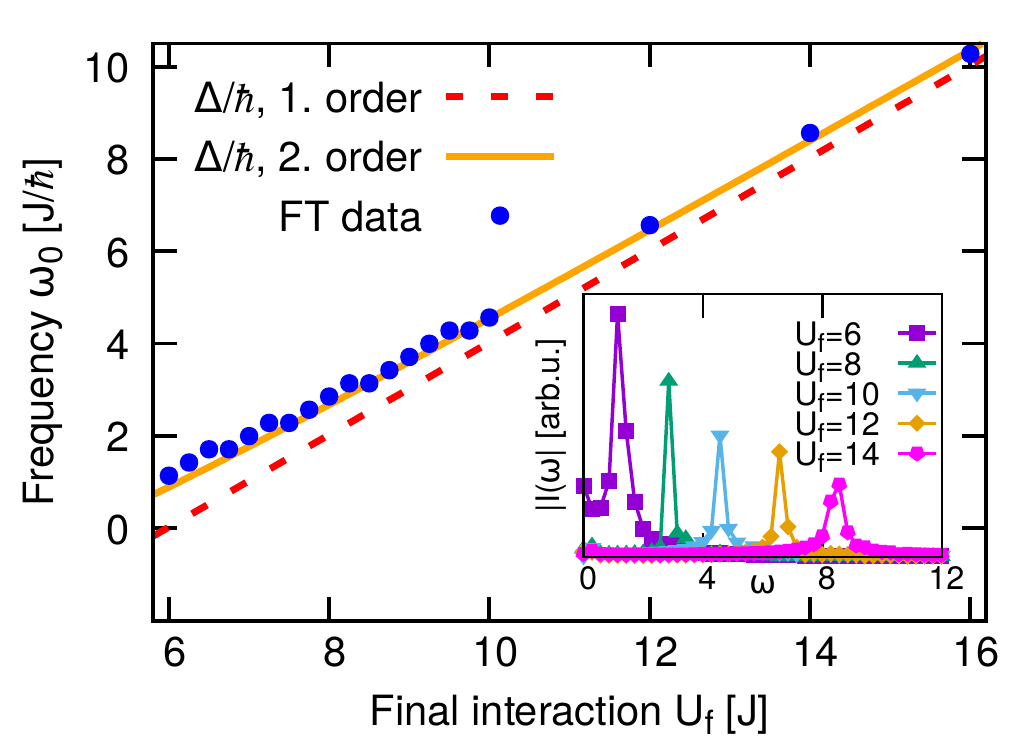}
	\caption{Position of the highest peak of the Fourier transform (FT, see inset) of the current for annealing rate $\hbar\gamma/J=1/6$ and field $\phi=0.7\pi$ in a 
	32-sites system (blue dots). 
	Dashed and solid curves show $\Delta(U)$ obtained from 
	strong-coupling perturbation theory up to $1^{\rm st}$ and $2^{\rm nd}$ order, see Eq.~\eqref{eqn:omega_0}.}
	\label{fig:ftmax_ttn}
\end{figure}

We now concentrate on the oscillation frequency, and in particular on the most relevant frequency, obtained by calculating the Fourier transform of $I(t)$ in an appropriate time window $[t_1,t_2]$
$I(\omega) = \int_{t_1}^{t_2} \! \ud t \, I(t) e^{-i\omega t} \;,$ where again $t_1> t_0$ as above in order to allow for a steady-state behavior to set in, and $t_2$ is the total 
simulation time. 
The Fourier transform of the current  $I(\omega)$ displays a sharp carrier frequency $\omega_0$, where $| I(\omega_0)|$ is maximum (see inset in Fig.~\ref{fig:ftmax_ttn}). 
The oscillation frequency $\omega_0$ is basically insensitive to the external flux $\phi$ and to the annealing rate $\gamma$. 
However, it does depend on the final on-site interaction $U_{\fin}$: it increases with $U_\fin$, as shown in Fig.~\ref{fig:ftmax_ttn}. 
We will now argue that, in the quasi-adiabatic limit, $\omega_0$ is essentially related to the gap $\Delta$ between the ground and the first excited state,
hence depends only on the final Hamiltonian.
A theoretical analysis using strong coupling perturbation theory for the ground and first excited state -- presented with more details in Sec.~\ref{sec:perturbation_} -- yields: 
\begin{equation} \label{eqn:omega_0}
\hbar\Delta (U_{\fin}) \simeq U_{\fin} - 5.97 J + 5.20\frac{J^2}{U_{\fin} } \;.
\end{equation}
The corresponding theory curves including up to first and second order terms are plotted in Fig.~\ref{fig:ftmax_ttn}, showing excellent agreement with the numerical results 
obtained from the Fourier transform of the current, hence confirming that $\hbar\omega_0=\Delta$. Furthermore we observe that the created defects  are not single holon-doublon pairs only, as in this case zeroth order perturbation theory -- taking into account only a single holon-doublon exicitation -- would fit the numerical data. The necessity of going to second order shows that the defects consist of multiple holon-doublon pairs, sites with more than two particles, and several empty sites.

\begin{figure}[t]
	\includegraphics[width=8.7cm]{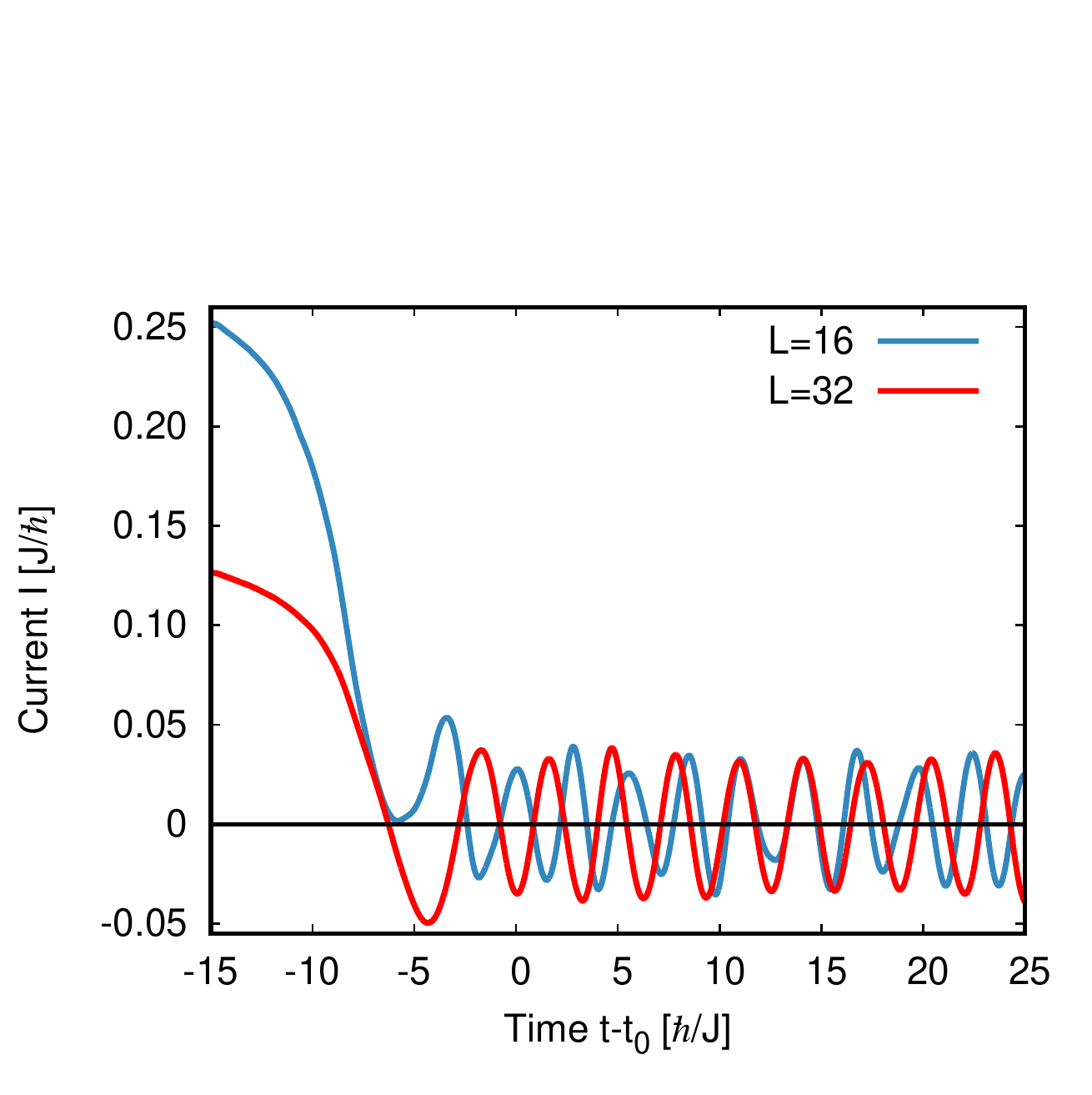}
	\caption{Dynamics of the current for two different system sizes $L=16$ and $32$ at annealing rate $\hbar\gamma/J=1/6$, magnetic field $\phi=0.7\pi$, and annealing from $U_{\ini}=2J$ to $U_{\fin}=7J$. }
	\label{fig:cur_difL}
\end{figure}

Let us now discuss the system size dependence of the results. First it is important to note that the equilibrium current of a perfect superfluid (interaction $U=0$) decreases as $1/L$ with the system size and therefore vanishes in the thermodynamic limit. For the oscillation amplitude of the current we can expect to find a similar behavior as long as the annealing is quasi-adiabatic, being the regime where we can relate the current amplitude to the residual energy. In Fig.~\ref{fig:res_energy_gamma_ttn} this behavior is indicated by the larger prefactor for $L=32$ as compared to $L=16$. On the other hand the energy gap at the critical point decreases with the system size (see Ref.\cite{PRB_Dziarmaga_2017}), leading to an increased population of excited states and correspondingly larger oscillations of the current. This effect is shown in Fig.~\ref{fig:cur_difL}, where the amplitudes for $N=16$ and $N=32$ are very similar for fixed annealing velocity, even though the initial values differ by a factor of $2$. However, keeping the annealing velocity fixed and increasing the system size will bring the dynamics out of the quasi-adiabatic regime. Therefore, if we stay in the quasi-adiabatic regime, the oscillations will vanish once we increase the system size.

\section{Strong coupling perturbation theory} \label{sec:perturbation_}
In the previous section we compared numerical results for the oscillation frequency of the current with the spectral gap at the end of the evolution as obtained from perturbation theory. In the following we provide some details of our perturbative calculations for the gap $\Delta$ in the strong repulsion limit. To this end we expand the energies of both, the ground state and the first excited state --- compatible with the symmetry --- up to second order. 
From here on we focus on the case $L=32$, while the field is fixed to be $\phi=0.7\pi$ as before.

It is important to note that the Bose-Hubbard model is translationally invariant, implying that the initial state of the dynamics -- a superfluid ground state -- is an eigenstate of the translation operator $\hat{T}$, where we find the corresponding eigenvalue to be 1. Due to the translational invariance of the model this eigenvalue is conserved, i.e. the time-dependent state will be an eigenstate of $\hat{T}$ with unit eigenvalue at any time. Later we will use this to construct a suitable basis. 

\subsection{Zeroth and first order}

In the infinite interaction limit $J/U = 0$ the ground state of Eq.~\eqref{eqn:HsysF} is a Mott insulator with one particle per site $|\text{GS}\rangle=|11...11\rangle$, such that to zeroth order the energy for the ground state is $E_0^{(0)}=0$. The first excited energy level is degenerate with $E_1^{(0)}=U$ for all states 
with one doublon and one holon.

Since the ground state is non-degenerate, the first order contribution to the ground state energy is given by 
\begin{equation*}
E_{0}^{(1)} = J \,\langle \text{GS}|\hat{V}|\text{GS}\rangle = 0,
\end{equation*}
where $\hat{V}=-\sum_j \left( e^{i\phi/L }b_{j+1}^{\dagger}b_{j} + H.c. \right)$ is the perturbation. 
In order to calculate the first order contribution to the excited states we use degenerate perturbation theory. 
First, we construct a suitable basis, where in particular we restrict the Hilbert space to the subspace of eigenstates of the translation operator $\Ttrans$ with unit eigenvalue. Let us define the basis states
\begin{equation*}
|s\rangle = \frac{1}{\sqrt{L}} \sum\limits_{q=0}^{L-1}(\Ttrans)^{q}|0\underbrace{11..11}_{\text{length} \ s}21..1\rangle
\end{equation*}
with $s=0,..,L-2$ the separation of the doubly occupied and the empty site. In this basis the perturbation matrix elements evaluate to
\begin{equation*}
\langle s'|\hat{V}|s\rangle = -3(\delta_{s,s'+1}e^{-i\phi/L} +\delta_{s',s+1}e^{i\phi/L} ).
\end{equation*}
The first order correction is given by the smallest eigenvalue of this $(L-1)\times (L-1)$ matrix, and is found numerically to be $E_{1}^{(1)} = - 5.97 J $ for $L=32$ and $\phi=0.7\pi$. Note that the exact eigenvalues of this tridiagonal Toeplitz matrix are known. However, we restrict ourselves to the numerical values for convenience in the next section.

\subsection{Second order}
The second order contribution for the ground state is given by
\begin{align}
E_0^{(2)} = (J^2/U) \sum\limits_{k\ne GS} \frac{|\langle k|\hat{V}|\text{GS}\rangle|^2}{E^{(0)}_0-E^{(0)}_k },
\end{align}
where $|k\rangle$ are the energy eigenstates for $J/U=0$. In this sum only states $|k\rangle$ with one doubly occupied and one neighboring empty site can give a nonzero contribution. Since the operator $\hat{V}$ contains $2L$ operators contributing $-2$ each, we find for the second order correction of the ground state energy:
\begin{align}
E_0^{(2)} =  (-4L)(J^2/U)
\end{align}

Similarly we continue for the second order contribution of the excited state, which is
\begin{align} 
E_1^{(2)} = (J^2/U) \sum\limits_{k \notin \{s\} } \frac{|\langle k|\hat{V}|1ex\rangle|^2}{E^{(0)}_1-E^{(0)}_k },
\end{align}
where $|1ex\rangle$ (not to be confused with $|s=1\rangle$) is the first excited state we found by diagonalizing the perturbation matrix. Using the numerical result for this state we can evaluate the sum to obtain the energy correction ($L=32$)
\begin{align}
E_1^{(2)} = -122.8(J^2/U).
\end{align}
Adding up all the different contributions we find for the gap:
\begin{equation} 
\hbar\Delta=E_1-E_0 \simeq U - 5.97 J + 5.20(J^2/U).
\end{equation}

%---------------------------------
\section{Discussion and Conclusion} \label{sec:conclusions}
%---------------------------------
%
Our  predictions are ready for experimental verification in future setups of cold atomic systems on ring lattices, where the Bose-Hubbard model including a 	gauge field can be naturally implemented with $^{87}$Rb atoms. Quasi-local currents can be measured following the read-out scheme presented in 
Ref.~\onlinecite{Nature_Trotzky_2012}. Due to the quasi-adiabatic annealing, the evolution times required for the observation of oscillations need to be 
longer than in commonly used sudden quench scenarios. However, the time periods in the order of $10\hbar/J$ needed here are still reachable in state-of-the-art experiments~[\onlinecite{Nature_Trotzky_2012}]. 
Moreover the non-decaying oscillations might eventually be used to demonstrate long-living coherence in next generation quantum simulators.

In conclusion, we have investigated the time-dependent behavior of the persistent current following a linear annealing procedure on a  Bose-Hubbard ring of up to $L=32$ sites, 
where in particular we analyzed the crossover from a superfluid to a Mott insulator in the presence of a gauge field. We found that after an initial decay the current 
starts to oscillate around nearly zero current, being the ground state value at the end of the ramp. The current is nonzero  due to the excitation of 
higher states and, in particular, results from the non-diagonal matrix elements of the current operator. In the slow annealing regime, where only one excited zero 
momentum state is occupied, the oscillation amplitude is proportional to the square root of the occupation probability and can therefore be related to the residual energy, characterizing the creation of defects. In a closed system the coherent oscillations are not expected to decay even for longer evolution times as long as the annealing is sufficiently slow. Instead, they persist according to Eq.~\eqref{eq:cur_decom}, due to the occupation of few eigenstates and correspondingly a small number of frequencies. This can easily be understood from the special case where only the ground and first excited state are populated. After the annealing ($t>t_0$) the populations stay constant, while the relative phase between the two populated eigenstates changes in time, producing the oscillations of the current. On the other hand, fast annealing leads to the presence of many frequencies, which might result in the decay of oscillations due to averaging effects. Using perturbation theory up to second order, we have been able to compute the frequency of the oscillations -- defined through the final Hamiltonian only -- in the 
limit of strong final interactions and slow annealing, in very good agreement with the numerical findings. While in this work we focused on the case where translational 
invariance is not broken, it might be interesting for future research to include a localized barrier, breaking the translational symmetry and providing the possibility to realize current-based qubits~\cite{Nature_Amico_2014,NJP_Aghamalyan_2015}. 

While finalizing this manuscript we became aware of the work by Bauernfeind and Aichhorn\cite{arxiv_Bauernfeind_2019}, presenting in detail the Time-dependent variational principle for general loopless Tensor Networks with an application to the Fork Tensor Product States tensor network used for Dynamical Mean-Field Theory calculations.

\section*{Acknowledgements} 
%We acknowledge fruitful discussions with ...... 
Research was partly supported by EU FP7 under ERC-ULTRADISS, Grant Agreement No. 834402,
the Quantum Flagship project PASQUANS, the EU via ERC synergy grant UQUAM and via QuantEra QTFlag, the Austrian Research Promotion Agency (FFG) via QFTE project AutomatiQ, the Italian PRIN 2017, the DFG funded TWITTER project,
and the US Air Force Office of Scientific Research (AFOSR) grant FA9550-18-1-0319. RF and GES acknowledge that their research has been conducted within the framework of the Trieste Institute for Theoretical Quantum Technologies (TQT).

\vspace{1cm}

\appendix
\section*{Appendix A: \\Numerical parameters/ convergence} \label{sec:convergence}
\begin{figure}[t]
	\centering
	\includegraphics[width=8.7cm]{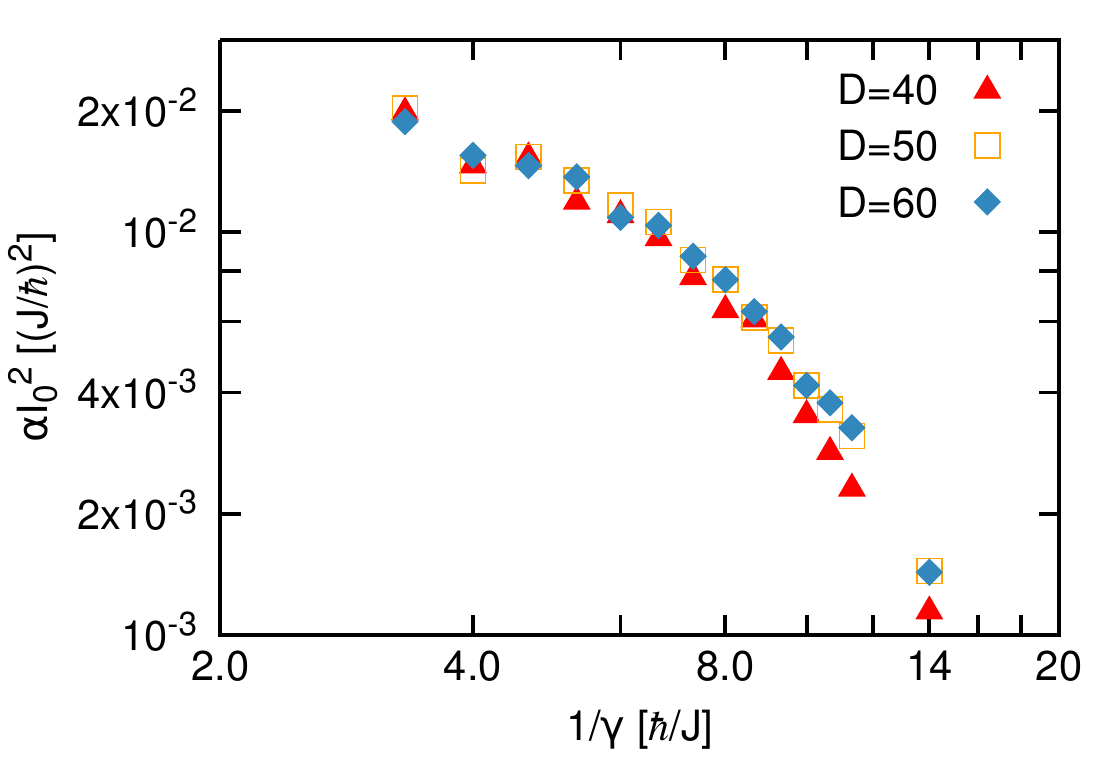}
	\caption{Square of the oscillation amplitude of the current $I_0^2$ multiplied by $\alpha=7.3$ as a function of the inverse annealing rate $1/\gamma$ for different bond dimensions $D$. While $D=50$ and $D=60$ curves agree well, significant differences appear for the bond dimension $D=40$. Physical parameters are those used in Fig.~\ref{fig:res_energy_gamma_ttn} and the system size is $L=32$.}
	\label{fig:bond_convergence}
\end{figure}

In this section we discuss the numerical parameters we used for our simulation. For the results shown previously we use the bond dimension $D=60$, the local bosonic dimension $d=5$ --- translating to the limitation of at most four particles per site ---, and the fixed time step $\Delta t=2\times 10^{-3}\hbar/J$. In the following we focus on the convergence in the bond dimension for the largest system considered ($L=32$), since numerical errors due to the truncation of the local boson occupation and the finite time step were found to be small compared to the error due to the bond dimension. As shown in Fig.~\ref{fig:bond_convergence} the oscillation amplitudes obtained for $D=60$ compare well with $D=50$, while more significant differences are visible in comparison with $D=40$. This indicates that indeed $D=60$ is enough to obtain accurate results. Note that in contrast to equilibrium scenarios, where the ground state energy decreases monotonically with the bond dimension, dynamical quantities like the oscillation amplitude can show non-monotonic behavior as a function of the bond dimension.

In practice it turned out that alternatively we can check if the bond dimension is sufficiently large by comparing the local currents between sites $k$ and $k+1$, defined as the time-dependent expectation value of the operator (compare with Eq.~\eqref{eqn:current_operator})
\begin{align}
\hat{I}_k 
= \frac{iJ}{\hbar}\left( e^{i\phi/L }b_{k+1}^{\dagger}b_{k} - H.c. \right) \label{eq:loc_cur}
\end{align}

\begin{figure}[h]
	\includegraphics[width=8cm]{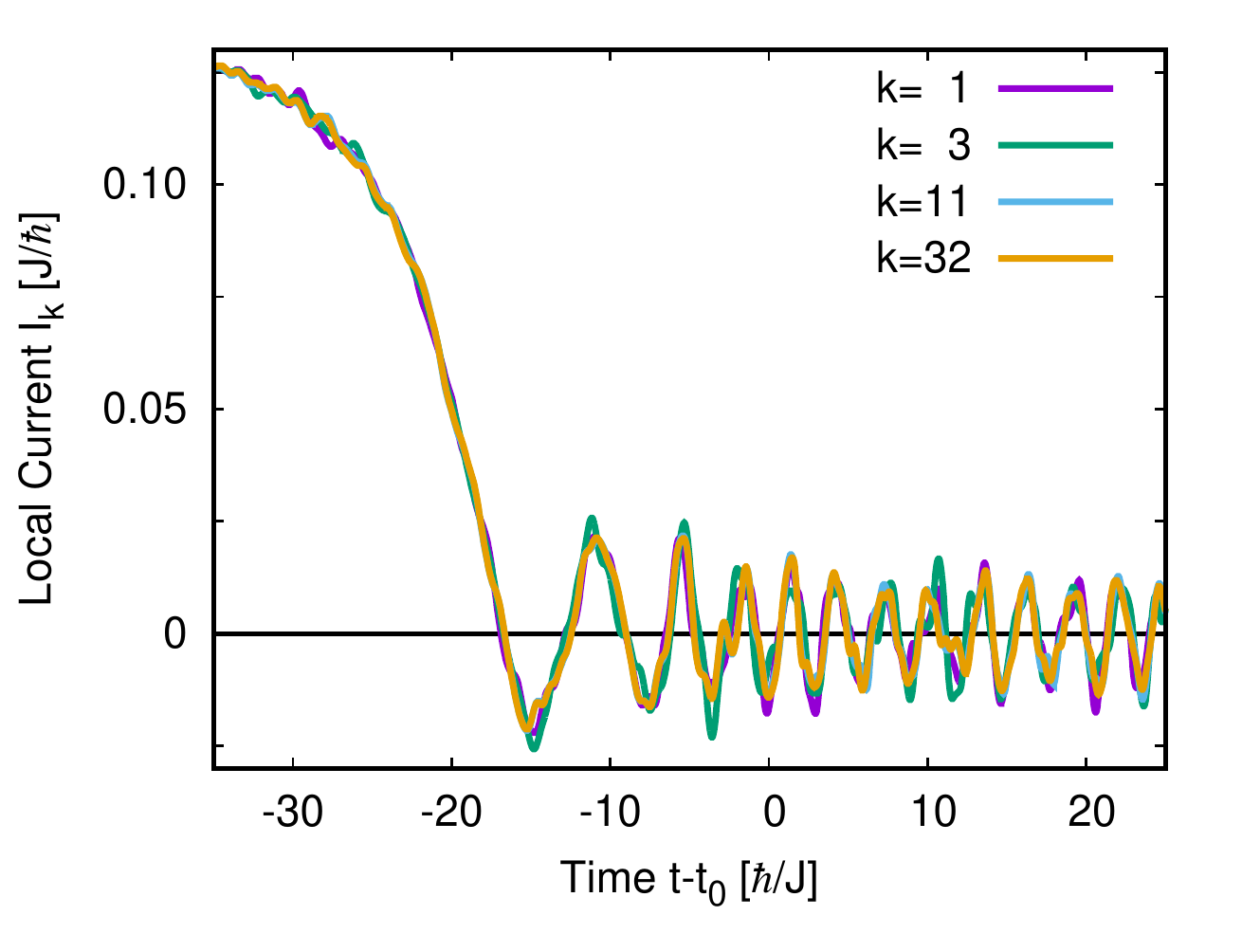}
	\includegraphics[width=8cm]{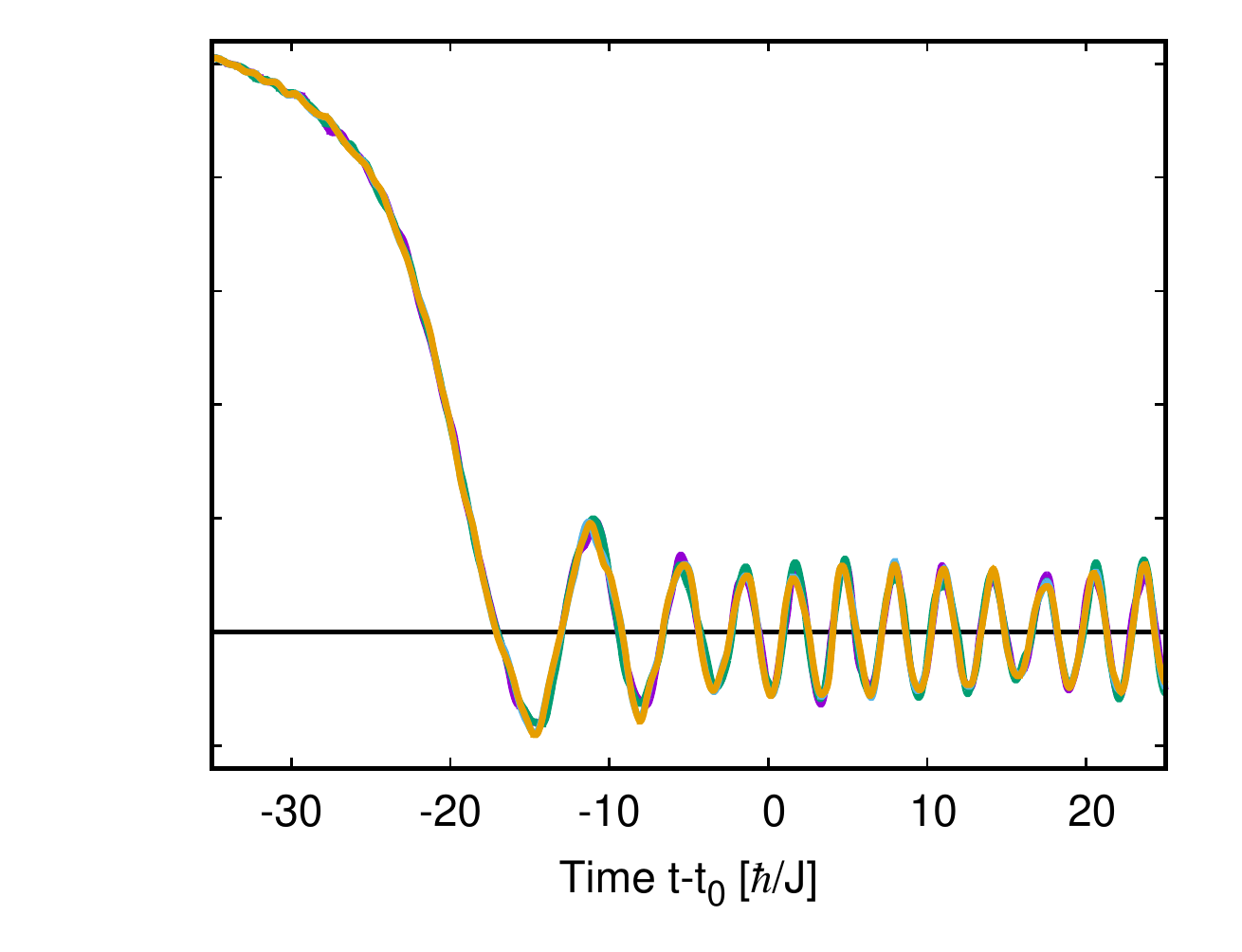}
	\begin{picture}(0,0)
	\put(-120,280){\Large $D=40$}
	\put(-120,120){\Large $D=60$}
	\end{picture}
	\caption{Dynamics of the local current $I_k$, given as the expectation value of the local current operator in Eq.~\eqref{eq:loc_cur}. The current between different neighboring sites $k$ and $k+1$ is shown for bond dimensions $D=40$ (top) and $D=60$ (bottom). Physical parameters are those used in Fig.~\ref{fig:cur_gamma_ttn}, at rate $\gamma\hbar/J=1/14$. }
	\label{fig:local_cur}
\end{figure}

Considering that the model of interest is translationally invariant, we expect to find the same local current between any pair of neighboring sites. However, the TTN breaks the translational invariance, resulting in different local currents if the bond dimension is too small. As visualized in Fig.~\ref{fig:local_cur}, the local currents do not agree very well for $D=40$, while the agreement is much better for $D=60$, confirming that $D=60$ yields reliable results.

\FloatBarrier

\bibliography{BiblioLNotes}

\end{document}